\RequirePackage{fix-cm}

\documentclass[twocolumn,epjc3]{svjour3}  
\usepackage[switch]{lineno}
\usepackage{subfig}
\usepackage{float}
\usepackage{caption}
\usepackage{ifthen}
\newboolean{uprightparticles}
\setboolean{uprightparticles}{false}
\usepackage{xspace}
\usepackage{upgreek}
\usepackage{xcolor}

\def\MC  {Monte\,Carlo simulation\xspace}

\def\cosTheta {\mbox{$\cos \Theta_{\mathrm{CS}}$}\xspace}
\def\ycm {\mbox{$y_{\mathrm{cm}}$}\xspace}
\def\data {\mbox{data}\xspace}
\def\muflux {\mbox{muon flux}\xspace}

\def\ux85 {\mbox{UX85}\xspace}

\ifthenelse{\boolean{uprightparticles}}%
{

 \def\Pmu         {\ensuremath{\upmu}\xspace}

 \def\Ppsi        {\ensuremath{\uppsi}\xspace}

 \def\PDelta      {\ensuremath{\Delta}\xspace}
 \def\PXi      {\ensuremath{\Xi}\xspace}
 \def\PLambda      {\ensuremath{\Lambda}\xspace}
 \def\PSigma      {\ensuremath{\Sigma}\xspace}
 \def\POmega      {\ensuremath{\Omega}\xspace}
 \def\PUpsilon      {\ensuremath{\Upsilon}\xspace}

 \def\PB      {\ensuremath{\mathrm{B}}\xspace}
 
 \def\PD      {\ensuremath{\mathrm{D}}\xspace}

 \def\PJ      {\ensuremath{\mathrm{J}}\xspace}
 \def\PK      {\ensuremath{\mathrm{K}}\xspace}

 \def\Pi      {\ensuremath{\mathrm{i}}\xspace}

}
{

 \def\Pmu         {\ensuremath{\mu}\xspace}

 \def\Ppsi        {\ensuremath{\psi}\xspace}
 
 \mathchardef\PDelta="7101
 \mathchardef\PXi="7104
 \mathchardef\PLambda="7103
 \mathchardef\PSigma="7106
 \mathchardef\POmega="710A
 \mathchardef\PUpsilon="7107
 
 \def\PB      {\ensuremath{B}\xspace}
 
 \def\PD      {\ensuremath{D}\xspace}

 \def\PJ      {\ensuremath{J}\xspace}
 \def\PK      {\ensuremath{K}\xspace}

 \def\Pi      {\ensuremath{i}\xspace}

}

\def\mumu       {\ensuremath{\Pmu^+\Pmu^-}\xspace}

\def\kaon  {\ensuremath{\PK}\xspace}
  \def\Kbar  {\kern 0.2em\overline{\kern -0.2em \PK}{}\xspace}

\def\Kz    {\ensuremath{\kaon^0}\xspace}
\def\Kzb   {\ensuremath{\Kbar^0}\xspace}
\def\KzKzb {\ensuremath{\Kz \kern -0.16em \Kzb}\xspace}
\def\Kp    {\ensuremath{\kaon^+}\xspace}
\def\Km    {\ensuremath{\kaon^-}\xspace}

\def\KpKm  {\ensuremath{\Kp \kern -0.16em \Km}\xspace}

  \def\Dbar    {\kern 0.2em\overline{\kern -0.2em \PD}{}\xspace}
\def\D       {\ensuremath{\PD}\xspace}

\def\Dz      {\ensuremath{\D^0}\xspace}
\def\Dzb     {\ensuremath{\Dbar^0}\xspace}
\def\DzDzb   {\ensuremath{\Dz {\kern -0.16em \Dzb}}\xspace}
\def\Dp      {\ensuremath{\D^+}\xspace}
\def\Dm      {\ensuremath{\D^-}\xspace}

\def\DpDm    {\ensuremath{\Dp {\kern -0.16em \Dm}}\xspace}

\def\Bbar    {\ensuremath{\kern 0.18em\overline{\kern -0.18em \PB}{}}\xspace}

\def\jpsi     {\ensuremath{{\PJ\mskip -3mu/\mskip -2mu\Ppsi\mskip 2mu}}\xspace}
\def\psitwos  {\ensuremath{\Ppsi{(2S)}}\xspace}

  \def\Y#1S{\ensuremath{\PUpsilon{(#1S)}}\xspace}

\def\Lbar {\ensuremath{\kern 0.1em\overline{\kern -0.1em\PLambda}}\xspace}

\def\AT#1     {\ensuremath{A_{\mathrm{T}}^{#1}}\xspace}           

\def\C#1      {\ensuremath{\mathcal{C}_{#1}}\xspace}                       
\def\Cp#1     {\ensuremath{\mathcal{C}_{#1}^{'}}\xspace}                    
\def\Ceff#1   {\ensuremath{\mathcal{C}_{#1}^{\mathrm{(eff)}}}\xspace}        
\def\Cpeff#1  {\ensuremath{\mathcal{C}_{#1}^{'\mathrm{(eff)}}}\xspace}       
\def\Ope#1    {\ensuremath{\mathcal{O}_{#1}}\xspace}                       
\def\Opep#1   {\ensuremath{\mathcal{O}_{#1}^{'}}\xspace}                    

\newcommand{\tev}{\ifthenelse{\boolean{inbibliography}}{\ensuremath{~T\kern -0.05em eV}\xspace}{\ensuremath{\mathrm{\,Te\kern -0.1em V}}\xspace}}
\newcommand{\gev}{\ensuremath{\mathrm{\,Ge\kern -0.1em V}}\xspace}
\newcommand{\mev}{\ensuremath{\mathrm{\,Me\kern -0.1em V}}\xspace}
\newcommand{\kev}{\ensuremath{\mathrm{\,ke\kern -0.1em V}}\xspace}
\newcommand{\ev}{\ensuremath{\mathrm{\,e\kern -0.1em V}}\xspace}
\newcommand{\gevc}{\ensuremath{{\mathrm{\,Ge\kern -0.1em V\!/}c}}\xspace}
\newcommand{\mevc}{\ensuremath{{\mathrm{\,Me\kern -0.1em V\!/}c}}\xspace}
\newcommand{\gevcc}{\ensuremath{{\mathrm{\,Ge\kern -0.1em V\!/}c^2}}\xspace}
\newcommand{\gevgevcccc}{\ensuremath{{\mathrm{\,Ge\kern -0.1em V^2\!/}c^4}}\xspace}
\newcommand{\mevcc}{\ensuremath{{\mathrm{\,Me\kern -0.1em V\!/}c^2}}\xspace}

\def\gsim{{~\raise.15em\hbox{$>$}\kern-.85em
          \lower.35em\hbox{$\sim$}~}\xspace}
\def\lsim{{~\raise.15em\hbox{$<$}\kern-.85em
          \lower.35em\hbox{$\sim$}~}\xspace}

\def\pt         {\mbox{$p_{\rm T}$}\xspace}

\def\tell1  {TELL1\xspace}
\def\ukl1   {UKL1\xspace}

\usepackage{orcidlink}
\usepackage{comment}
\usepackage{nicematrix}

\smartqed  
\RequirePackage{graphicx}
\journalname{Eur. Phys. J. C}
\begin{document}

\title{Measurement of di-muons from 400 GeV/c protons interacting
in a thick molybdenum/tungsten target
}


\author{The SHiP Collaboration\thanksref{e1,addr1}
}

\thankstext{e1}{e-mail: thomas.ruf@cern.ch}


\institute{The full authorlist is supplied at the back of this article. \label{addr1}
}

\date{Received: date / Accepted: date}

\maketitle

\begin{abstract}
Di-muon events emanating from a replica of the SHiP target  at the CERN SPS (with a 400 GeV/c proton beam) contain 
a clear signal of \jpsi production. The target consisted of 13 interaction lengths of molybdenum and tungsten, followed by a 2.4m iron hadron absorber. The di-muon rate is in reasonable
agreement with \MC based on Pythia v8. However, the $\jpsi$ rate observed in the data is much lower than predicted by the Monte Carlo simulation, which is advantageous for SHiP since  muons with a large $\pt$ are more difficult to deflect by the active muon shield.

For the center of mass rapidity ($\ycm$) interval with the largest overlap with the NA50 measurement, $0.3 < \ycm < 0.6$, we obtain the production cross-section per nucleon $\sigma(\jpsi)/A$ including the branching ratios of $\jpsi$ decays into muons, $B_{\mumu}$:

$B_{\mumu}\sigma(\jpsi)/A=(1.18~\pm~0.04~\pm~0.10)~\mathrm{nb}$/nucleon.

This is to be compared to the NA50 extrapolated result (using a much thinner target):
 
$B_{\mumu}\sigma(\jpsi)/A=(0.99~\pm~0.04)~\mathrm{nb}$/nucleon. 

Within the systematic errors, no significant enhancement due to secondary production of \jpsi inside the target is observed. An upper limit at 90\% confidence level of possible contributions from secondary collisions of $<32\%$ is obtained.

\keywords{\jpsi \and di-muon \and cross section \and SHiP}
\end{abstract}
\section{Introduction}
\label{sec:intro}

For the optimization of the muon shield magnets of the SHiP\footnote{Approved in April 2024 as CERN experiment NA67.} experiment~\cite{SHiP}, a good understanding of the production of muons that can lead to signal-like events is mandatory. During a 4 week test beam period in 2018 the momentum spectrum of the muons emanating from a SHiP-like heavy target was measured at the CERN SPS with $400\,\gevc$ protons. This measurement validated the expectations for single muons from our Monte Carlo simulation~\cite{muflux}. This reference contains the details of the experimental setup, data acquisition and analysis. In the current paper we report the observation of a \jpsi signal in the 2018 data using events with two or more reconstructed tracks.  \jpsi mesons produced in the interactions, and subsequently decaying to two muons, are expected to be a major contributing source of high $\pt$ muons which are a significant background for SHiP. 

The NA50 experiment~\cite{NA50:2003pvd,NA50:2003fvu} measured the \jpsi cross section for $400\,\gev$ protons impinging on several thin targets~\cite{Alessandro:2006jt} made of Be, Al, Cu, Ag, W and Pb.
Comparing our measurement with the 1.5\,m-long SHiP target to the results of NA50 we can estimate the contribution of the secondary production.

This paper is organized as follows. A summary of the data and \MC samples is given in Section~\ref{sec:DataMC}, followed by the track reconstruction and dedicated corrections in Section~\ref{sec:trackReco}. The \jpsi di-muon selection is detailed in Section~\ref{sec:recoJpsi} leading to the extraction of differential cross sections as a function of rapidity and \cosTheta (the polar angle in the Collins-Soper reference frame~\cite{Collins:1977iv}) in Section~\ref{sec:xsec}. A summary is given in Section~\ref{sec:summary}.
\section{Data and MC samples}
\label{sec:DataMC}
The 2018 muon flux measurement setup consisted of a target of $1.5~$m of target material (13 interaction lengths of molybdenum and tungsten followed by a $2.4~$m iron hadron absorber), a spectrometer tracker composed of four drift-tube stations and a magnet (deflection in the horizontal plane), and a muon tagger made of resistive plate chambers (RPCs). The details of this setup can be found in Ref.~\cite{muflux}.

The right handed coordinate system is defined as follows: the positive $z$ axis follows the beam from the target and the positive $y$ axis points vertically upwards. The center of mass rapidity \ycm is defined as: 
\begin{equation}
\ycm = \frac{1}{2} \ln \frac{E+p_z}{E-p_z} - y_{\mathrm{beam}}, y_{\mathrm{beam}}\approx \frac{1}{2} \ln \frac{2E_{\mathrm{beam}}}{M_N}
\end{equation}
with $E_{\mathrm{beam}}$ equal to the beam energy, $M_N$ the average nucleon mass and $E$ and $p_z$ the energy of the particle and momentum in the beam direction in the laboratory frame respectively. For $400\,\gevc$ protons, $y_{\mathrm{beam}} = 3.374$. The center of mass pseudorapidity is defined as $\eta_{\mathrm{cm}} = \eta - y_{\mathrm{beam}}$. The geometric acceptance, defined by the polar angle  with respect to the beam, was $\Theta_\mu = 0.054$ rad. This corresponds to a pseudorapidity  of $\eta = 3.6$, or  $\eta_{\mathrm{cm}} = 0.2$.

The beam intensities  were less than  $3\times 10^6$ protons per second and the data set corresponded to $(3.25 \pm 0.07) \times 10^{11}$ protons on target (POT). Only one out of  $710\pm 15$ collisions~\cite{normalization} produced an event with at least one fully reconstructed track, resulting in a maximum rate of $\sim  4.2$ kHz. 
In our data we observe  about $0.05\times 10^{-3}$ same sign and $6.0\times 10^{-3}$ opposite sign events per POT with a fully reconstructed track. The excess of opposite sign events of two orders of magnitude is clearly due to associated production in one proton-nucleon collision.

For the analysis in Ref.~\cite{muflux}, muon events from $66.02 \times 10^9$ POT were produced using Pythia v6~\cite{Pythia6} and Pythia v8~\cite{Pythia8} (using the default settings) together with GEANT4 \cite{Geant4}. 
However, the fraction of protons which produce heavy flavour for a $400\gevc$ proton beam colliding on
a molybdenum target do not take into account that the target is several
interaction-lengths long.  Secondaries produced in the initial $pA$ collision can
produce heavy flavour in a subsequent interaction. The heavy flavour
production for both the primary and secondary interactions, and the corresponding phase-space
distribution of the hadrons were obtained by using Pythia v6.4. At the time of this production Pythia v8, unlike Pythia v6,  had to be reinitialized for each $pA$ collision,  resulting in an unaffordable computing overhead. Further technicalities and how these productions are subsequently mixed with the minimum bias from Pythia v8 are described in Ref.~\cite{cascade}.

\begin{figure}[ht]
\centering
\includegraphics[width=\columnwidth]{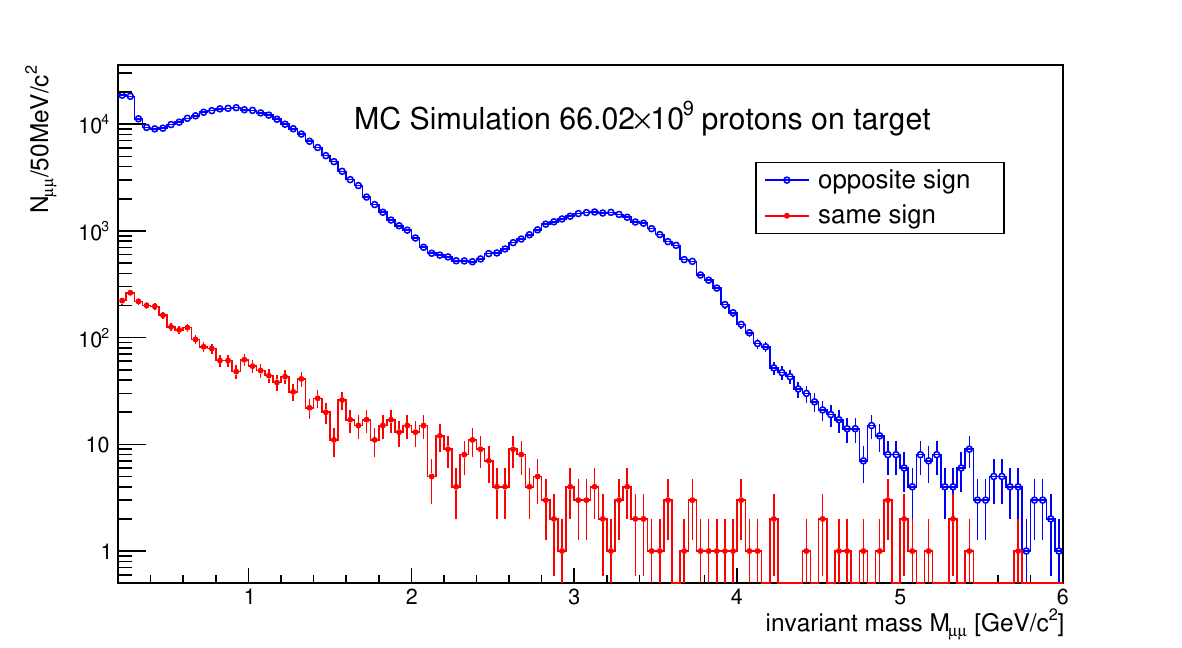}
\includegraphics[width=\columnwidth]{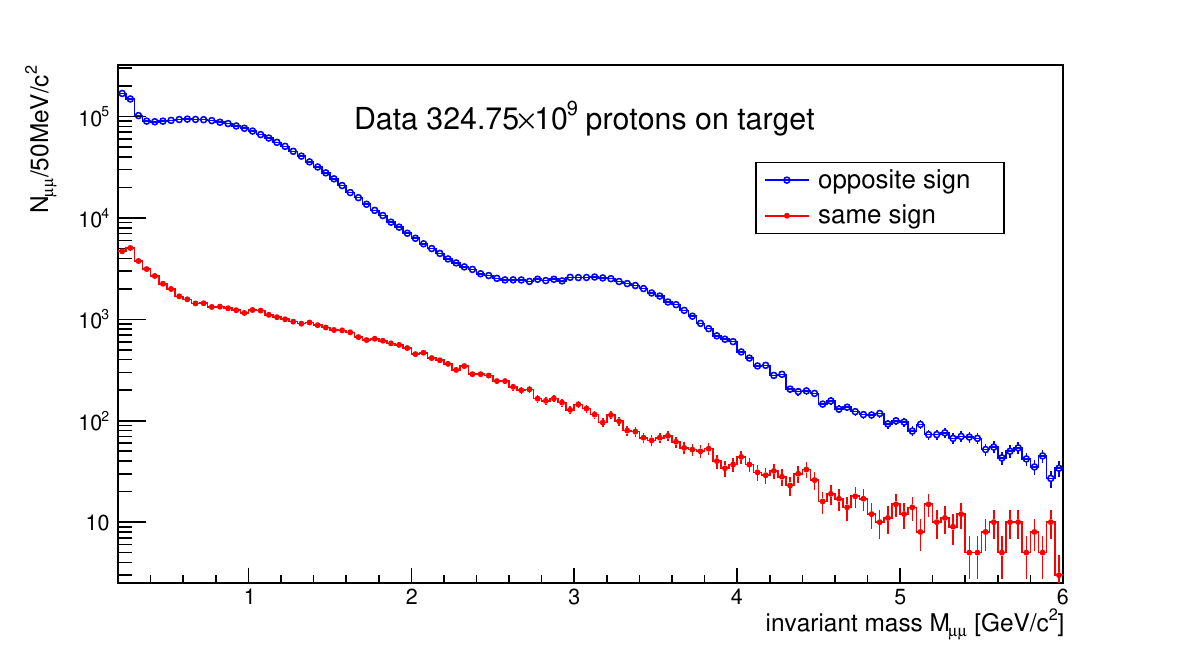}
\caption{Reconstructed invariant mass of opposite and same sign combinations, top \MC, bottom \data.}
\label{fig:OppAndSameSign}
\end{figure}

\begin{figure}[ht]
\centering
\includegraphics[width=\columnwidth]{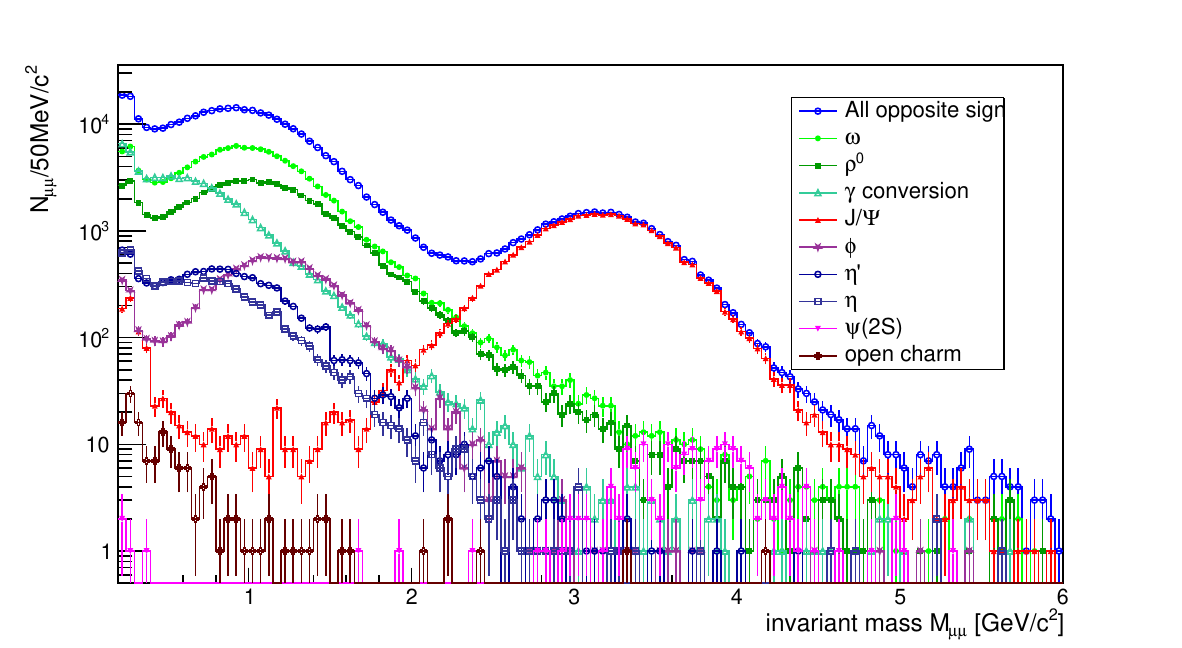}
\caption{Reconstructed invariant mass of opposite sign combinations in the \MC  for different sources.}
\label{fig:OriginMu}
\end{figure}

Since this \MC was produced with a $10\gev$ threshold on the kinetic energy, a minimum cut on the momenta of the two reconstructed tracks has to be applied. For $p>20\,\gev$ muon tracks, we observe $2.6 \times 10^{-2}$ same sign events  in \data  and $ 0.8 \times 10^{-2}$ in the Monte Carlo simulation (see Figure~\ref{fig:OppAndSameSign}). The contribution from same sign events is very small. They originate from random combinations and wrong charge assignments. The dominant sources for opposite sign events in the \MC are listed in Table~\ref{tab:OriginMu} and their reconstructed invariant mass is shown in Figure~\ref{fig:OriginMu}. 

The amplitudes of the bumps in Figure~\ref{fig:OppAndSameSign} are not compatible between Monte Carlo simulation and data. Pythia is known to predict charm production at low energies rather poorly (see Refs.~\cite{cascade,LOURENCO2006127}).  The low mass bump is produced by dimuon decays of low mass resonances, which are themselves produced in the primary proton nucleon collision, but also in subsequent secondary production at lower energies. This makes it a challenge for the Monte Carlo simulation to predict the precise number.  

\begin{table}[ht]
\centering
\caption{Main sources of di-muons in the \MC.}
\label{tab:OriginMu}
\begin{NiceTabular}{| l | l |}
 \hline
source & N / $10^6$ POT \\
\hline
            $\omega$ &   $1.93$ \\
            $\rho^0$   &  $1.03$ \\
            $\gamma$ conversion &    $0.889$ \\
            \jpsi &  $0.45$ \\
            $\phi$    &  $0.16$ \\
            $\eta'$     &  $ 0.14$ \\
            $\eta$    &  $0.11$ \\
            $\psi(2S)$   &  $ 0.002$ \\
             open  charm   &  $0.002$ \\ 
\hline
\end{NiceTabular}
\end{table}

Two different Monte Carlo generators have been used for \jpsi production. Pythia v8~\cite{Pythia8}, with only \jpsi production from primary collisions, and Pythia v6~\cite{Pythia6} for \jpsi production for primary and secondary collisions~\cite{cascade}.

\begin{figure}[ht]
\centering
\includegraphics[width=\columnwidth]{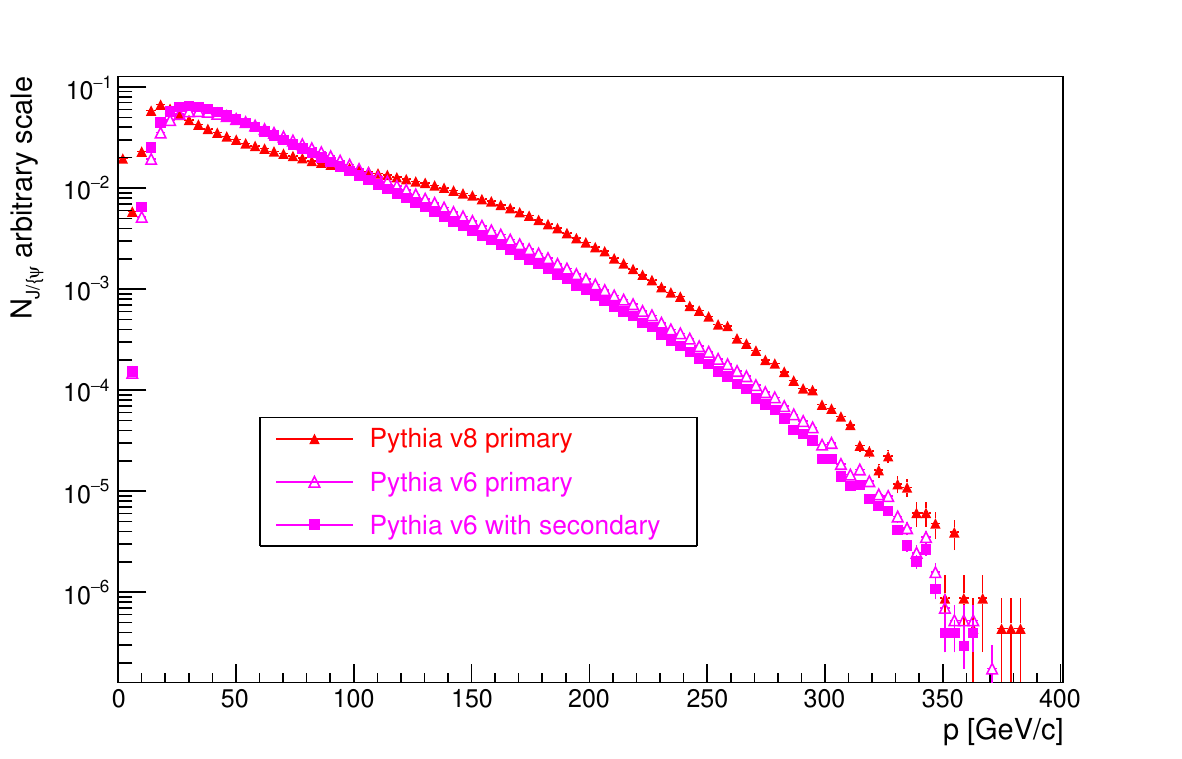}
\includegraphics[width=\columnwidth]{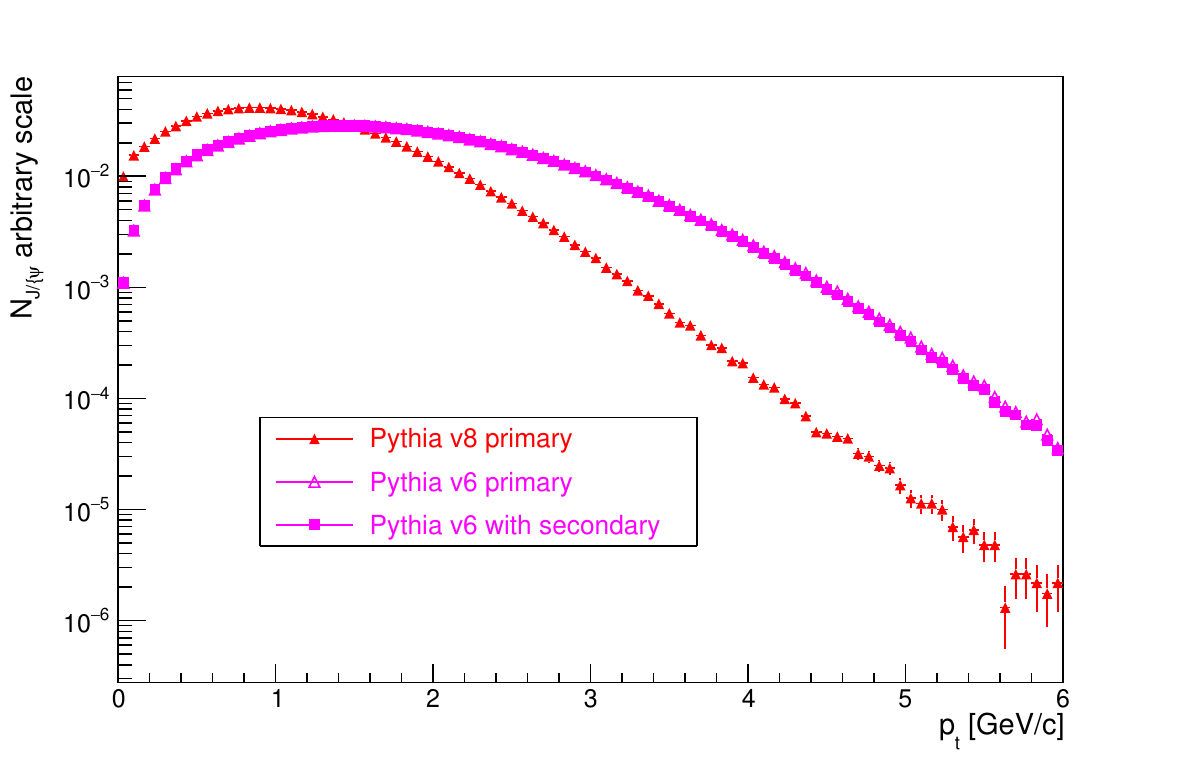}
\includegraphics[width=\columnwidth]{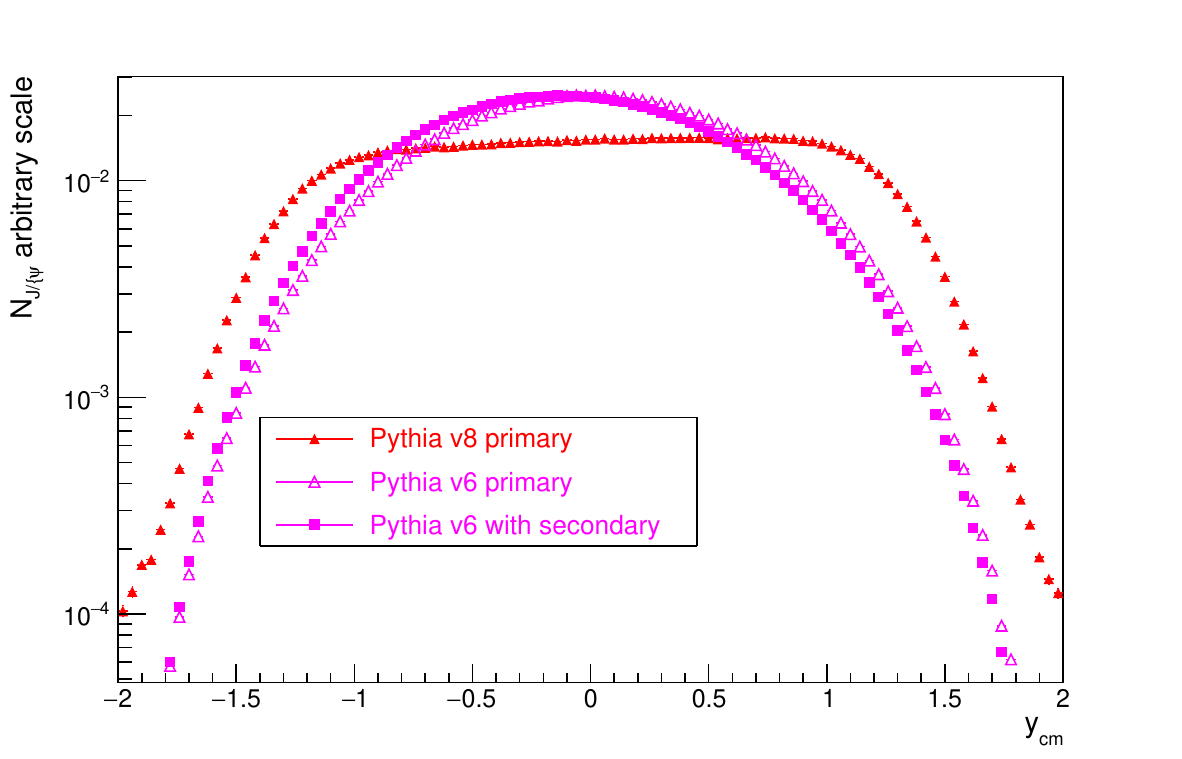}
\caption{Momentum, transverse momentum and \ycm of \jpsi in the \MC. The distributions are normalized to 1, in order to compare the shapes only. For Pythia v6, there is little difference between only primary and including secondary \jpsi production.}
\label{fig:JpsiKinematicsProj}
\end{figure}

The momentum, transverse momentum and \ycm distributions of the simulated \jpsi are shown in Figure~\ref{fig:JpsiKinematicsProj} for both generators. The transverse momentum distribution from Pythia v6 is much harder than from Pythia v8. In both cases, the transverse momentum distribution is independent of the rapidity. The rapidity distributions are also very different between Pythia v6 and Pythia v8.  The reason for this is not understood.

More details on the kinematics of the muons from \jpsi decays can be found in Ref.~\cite{SHiP:DiMuon}.
\section{Track reconstruction and energy-loss corrections}
\label{sec:trackReco}
In the \jpsi reconstruction, two effects have to be accounted for and corrected: the loss of energy of the muons in the material upstream of the spectrometer tracker, and the change in direction caused by multiple scattering between the \jpsi production point and the track momentum measurement in the spectrometer tracker. The muons travel through at most $1.5$~m of target material followed by $2.4~$m of iron before reaching the spectrometer. As determined by Geant4~\cite{Geant4}, they lose on average about $10\,\gev$ of energy.

The change in direction of the reconstructed momentum caused by multiple scattering is partly mitigated by replacing the momentum direction obtained from the track fit with the direction defined by the position of the centre of the target and the first measured point. Henceforth, when referring to momentum, we imply that these corrections were applied.
\section{\jpsi reconstruction and selection}
\label{sec:recoJpsi}

\begin{figure}[ht]
\centering
\includegraphics[width=\columnwidth]{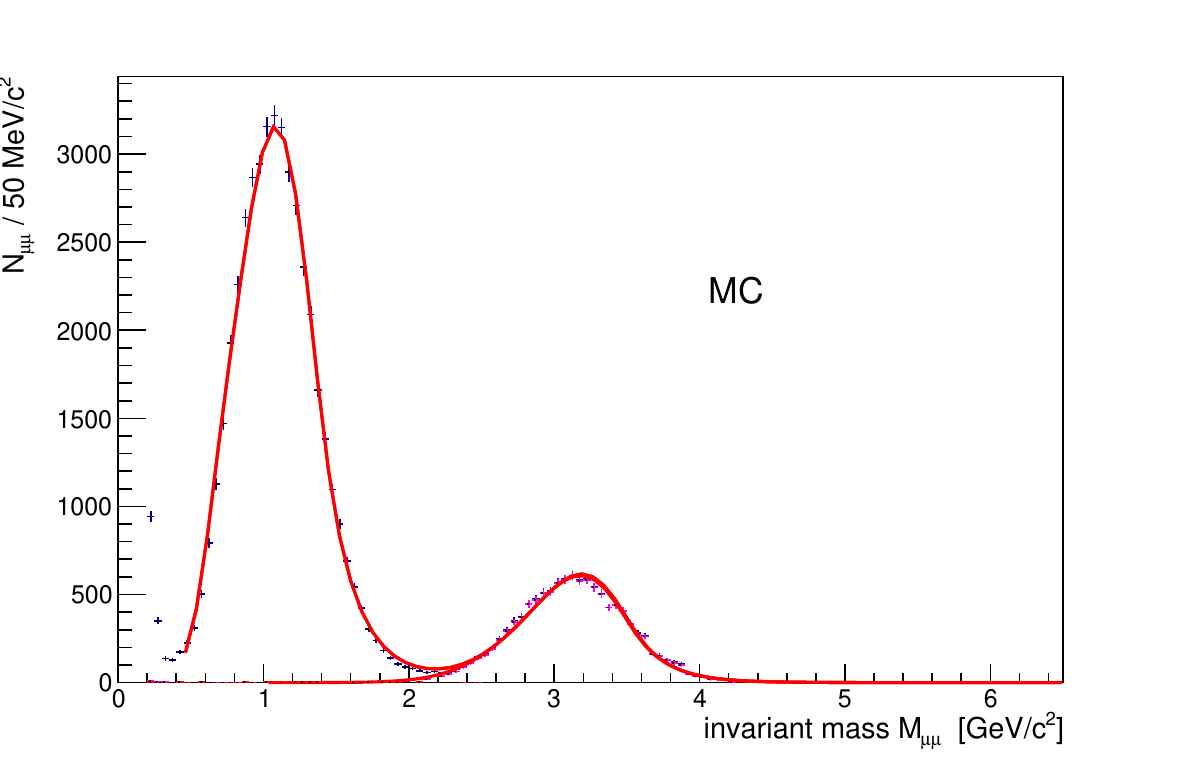}
\includegraphics[width=\columnwidth]{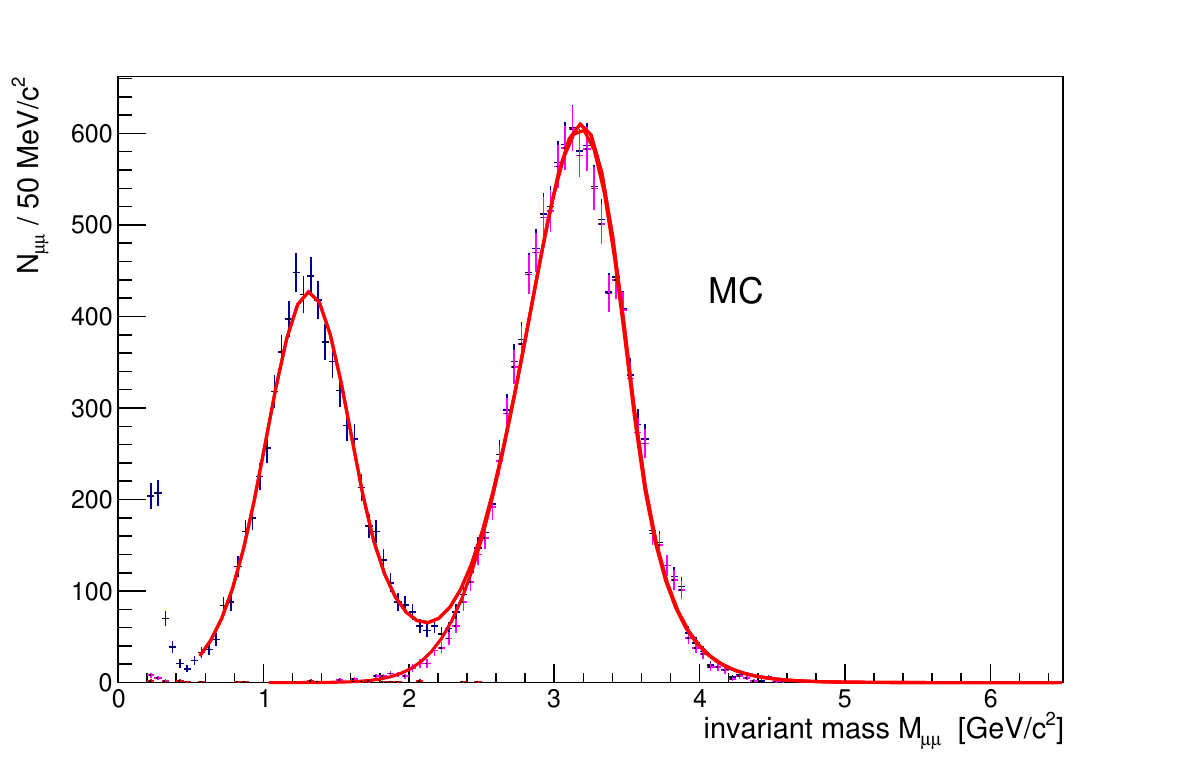}
\caption{Invariant mass of di-muon pairs in the  \MC. Top without minimum \pt selection, bottom with at least one track with $\pt>1.0\,\gevc$. The distributions are fitted with the Bukin model.}
\label{fig:InvMass}
\end{figure}

Candidate \jpsi events are required to contain two tracks with a minimum momentum of $>20\,\gevc$ and a maximum momentum of $<300\,\gevc$. The lower threshold is motivated by the large multiple scattering of low momentum tracks (see Section~\ref{sec:trackReco}). The upper threshold suppresses ghost tracks~\cite{muflux}. The invariant mass distribution of di-muon pairs in the \MC is shown in Figure~\ref{fig:InvMass}, top. The peak at very low mass is an artifact of the correction for multiple scattering, due to the finite size of a drift tube. If the first measured hit of both muon tracks is identical, the correction results in a zero opening angle and therefore the invariant mass becomes twice the muon mass. The contribution of same sign muon pairs is very low, and mainly located at low masses. The peak at $\sim 1~\gevcc$ is due to decays of low mass resonances ($\eta$, $\omega$, $\rho^0$, $\Phi$) and muon pair production in photon conversions, see also Table~\ref{tab:OriginMu}. The peak at higher mass is due to \jpsi decays. The contribution of \jpsi decays can be significantly enhanced compared to the low mass by requiring at least one track with $\pt>1.0\,\gevc$ as shown in the bottom plot of Figure~\ref{fig:InvMass}.

Due to the large energy loss and multiple scattering, the two track invariant mass distribution cannot be fitted well with two Gaussian distributions. The following fit models are used:
\begin{itemize}
\item one Crystal Ball function~\cite{Skwarnicki:1986xj} with a tail to higher invariant mass accounting for the low mass resonances and a second Crystal Ball function with a tail towards lower mass for the \jpsi signal.
\item one Bukin function~\cite{bukin}   with asymmetric tails towards low and high  masses for the low mass resonances and a second Bukin function for the \jpsi signal.
\end{itemize}

The Bukin fit results are taken as the baseline and deviations from the results using the Crystal Ball fit model are added to the systematic errors.

\begin{figure}[ht]
\centering
\includegraphics[width=\columnwidth]{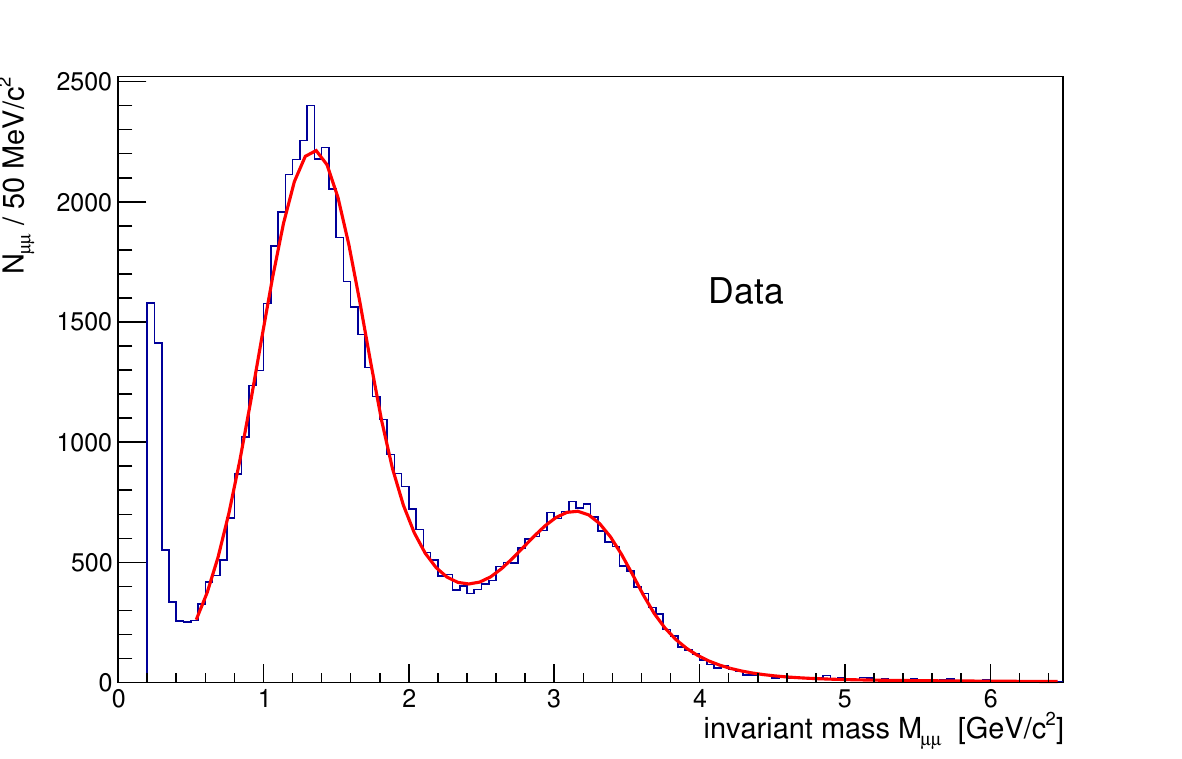}
\caption{Fitted di-muon invariant mass distribution in the data with a free \psitwos component and with at least one track with $\pt>1.0\,\gevc$.}
\label{fig:psi2s}
\end{figure}

The di-muon invariant mass distribution from data with at least one track with $\pt>1.0\,\gevc$ is shown in Figure~\ref{fig:psi2s}. It was fitted with a \psitwos contribution with the same shape parameters as for the \jpsi, only with a free normalization. The values of the fitted parameters for Monte Carlo simulation are shown in Tables~\ref{tab:Fig12a} and \ref{tab:Fig12b}. For data they are shown in Table~\ref{tab:Fig13}. 

The Monte Carlo simulation agrees well with the data yields for the contribution from low mass resonances after scaling with the different number of protons on target. The ratio data/Monte Carlo $\mathrm{R_{low\, mass}} = 1.03\pm 0.03_\mathrm{stat.}$, while the ratio for the \jpsi is  $\mathrm{R}_{\jpsi} = 0.22\pm 0.01_\mathrm{stat.}$.

\begin{figure}[h]
\centering
\includegraphics[width=\columnwidth]{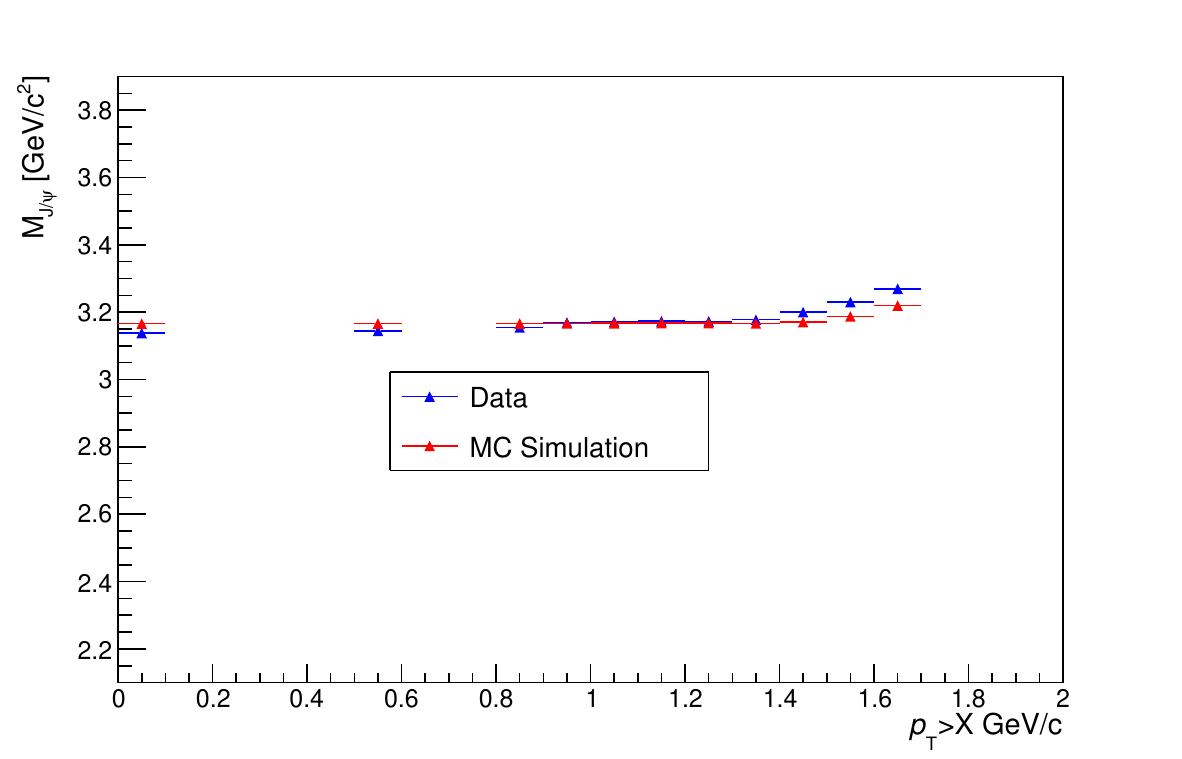}
\caption{Evolution of the fitted value of the invariant mass as function of minimum \pt threshold on at least one track for data (blue) and \MC(red).  The horizontal bars correspond to the bin width. }
\label{fig:InvMassEvol}
\end{figure}

It is important to note that the fitted \jpsi mass in data and Monte Carlo simulation are in good agreement (see also Figure~\ref{fig:InvMassEvol}). They do not agree with the PDG value due to the incomplete correction of the large energy loss. The energy loss is non-gaussian with large tails and the only correction which we
can apply is to correct for the mean energy loss. Due to the tails, this does not necessarily result in the correct \jpsi mass. One could in principle adjust the energy correction to reproduce the PDG value of the \jpsi mass, but we find this unnecessary. The deviations in the Monte Carlo simulation and data are similar.

\begin{table}[ht]
\centering
\caption{Fit results from the top plot of Figure~\ref{fig:InvMass} for Monte Carlo simulation without a minimum \pt selection. The distribution was fitted with the Bukin model.}
\label{tab:Fig12a}
\begin{NiceTabular}{| l | l |}
 \hline
Entries        &   $ 57563$ \\
$\chi^{2}$/ndf &   $   394 / 144$ \\
$\mathrm{Signal_{low\,mass}}$      &   $ 6.38\times 10^4 \pm 2.3\times 10^2$ \\
$\mathrm{Mean_{low\,mass}}$        &   $ 1.083 \pm 0.001$ \\
$\sigma_{\mathrm{low\,mass}}$       &   $ 0.2798 \pm 0.0006$ \\
\jpsi      &   $  1.24\times 10^4 \pm 91$ \\
Mass           &   $  3.191 \pm 0.003$ \\
$\sigma$          &   $  0.3414 \pm 0.0017$ \\
\hline
\end{NiceTabular}
\end{table} 

\begin{table}[ht]
\centering
\caption{Fit results from the bottom plot of Figure~\ref{fig:InvMass} for Monte Carlo simulation with at least one track with $\pt>1.0\,\gevc$. The distribution was fitted with the Bukin model.}
\label{tab:Fig12b}
\begin{NiceTabular}{| l | l |}
 \hline
Entries        &   $ 18610$ \\
$\chi^{2}$/ndf &   $   206.3 / 142$ \\
$\mathrm{Signal_{low\,mass}}$      &   $ 8603 \pm 142.0$ \\
$\mathrm{Mean_{low\,mass}}$         &   $ 1.313 \pm 0.004$ \\
$\sigma_{\mathrm{low\,mass}}$      &   $ 0.3013 \pm 0.0040 $ \\
\jpsi          &   $ 1.23\times 10^4 \pm 150$ \\
Mass           &   $  3.189 \pm 0.004$ \\
$\sigma$            &   $  0.348 \pm 0.003$ \\
\hline
\end{NiceTabular}
\end{table}

 A ratio of \psitwos over \jpsi of $(0.7\pm0.9)\%$ is obtained corresponding to an upper limit of $<2.1\%$ for the \psitwos contribution to the \jpsi production at $90\%$ confidence level. In comparison, the precise mass resolution of the NA50 measurement enabled the determination of the contribution of the $\psitwos$ and Drell-Yan production to the di-muon cross section, with a $\psitwos$ contribution of $1.5\%$ to $1.9\%$ depending on the target material. This is neglected in our fit and covered by the systematic error. The Drell-Yan contribution in the mass range $[2.9 - 4.5] \gevcc$ is of the same order and partially accounted for in the background contribution.

\begin{table}[ht]
\centering
\caption{Fit results with a free \psitwos component and with at least one track with $\pt>1.0\,\gevc$. The distribution was fitted with the Bukin model.}
\label{tab:Fig13}
\begin{NiceTabular}{| l | l |}
 \hline
Entries        &   $ 65497  $ \\
$\chi^{2}$/ndf &   $   360.5 / 136 $ \\
$\mathrm{Signal_{low\,mass}}$      &   $ 4.37\times 10^4 \pm 2.8\times 10^2$ \\
$\mathrm{Mean_{low\,mass}}$        &   $ 1.344 \pm 0.002 $ \\
$\sigma_{\mathrm{low\,mass}}$      &   $ 0.377 \pm 0.004  $ \\
\jpsi        &   $ 1.34\times 10^4 \pm 1.2\times 10^2$ \\
Mass           &   $  3.162 \pm 0.008$  \\
$\sigma$          &   $  0.429 \pm 0.007 $ \\
\psitwos       &   $  98 \pm 115 $ \\
\hline
\end{NiceTabular}
\end{table}

The ratio of signal yields from \MC over data as a function of the  minimum \pt threshold  stays constant within $\pm 5\%$ for a wide range of selections. The efficiency of this selection is therefore well reproduced with the \MC.  Figure~\ref{fig:InvMassEvol} shows the fitted value of the invariant mass as function of the  minimum \pt threshold. The \MC and data agree well with each other.

\begin{figure}[ht]
\centering
\includegraphics[width=\columnwidth]{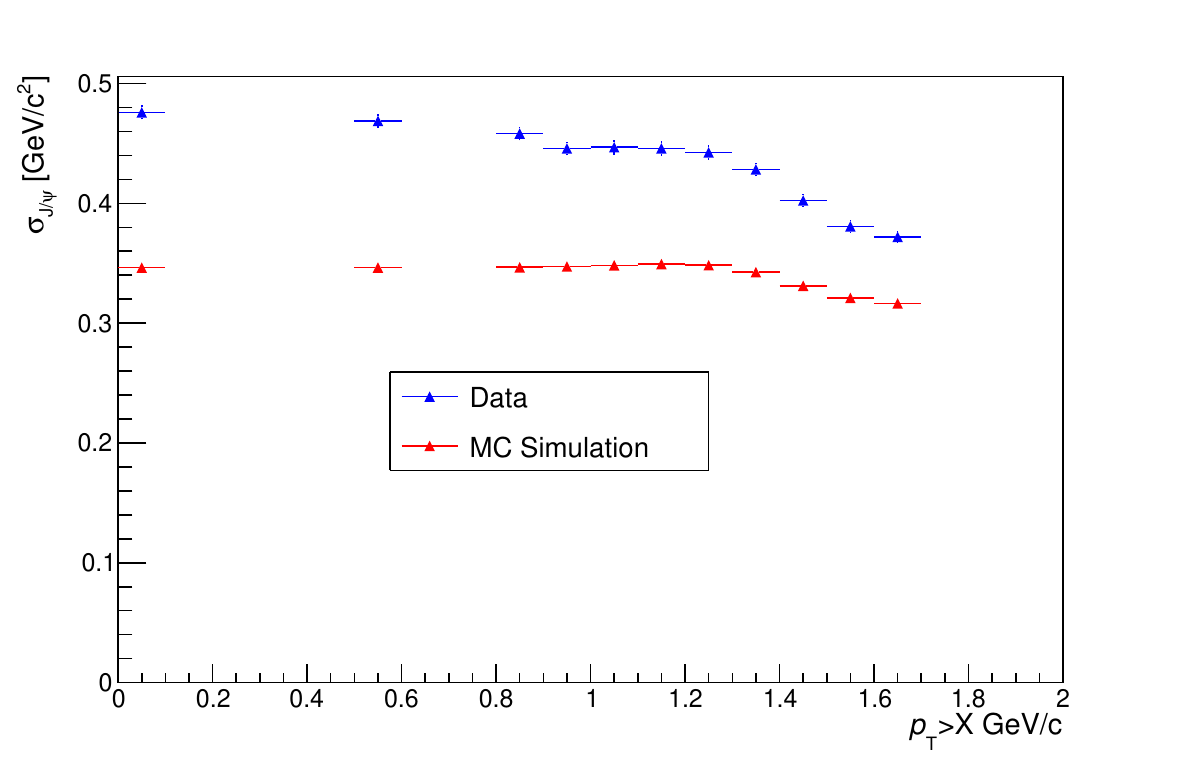}
\caption{Evolution of the invariant mass resolution as function of minimum \pt threshold on at least one track for data (blue) and \MC(red).}
\label{fig:InvSigmaEvol}
\end{figure}

Figure~\ref{fig:InvSigmaEvol} shows the invariant mass resolution as function of the minimum \pt threshold. The resolution in data is much worse compared to  the \MC. This could be explained by a larger spread of energy loss or larger multiple scattering in data than predicted by Geant4. The latter would be important to study further with regard to the active muon shield~\cite{mushield} for the SHiP experiment.

The yield of \jpsi as a function of reconstructed momentum, transverse momentum, rapidity, and \cosTheta is obtained  by fitting the di-muon invariant mass distributions in bins of the kinematic variable. 

\begin{figure}[ht]
\centering
\includegraphics[width=\columnwidth]{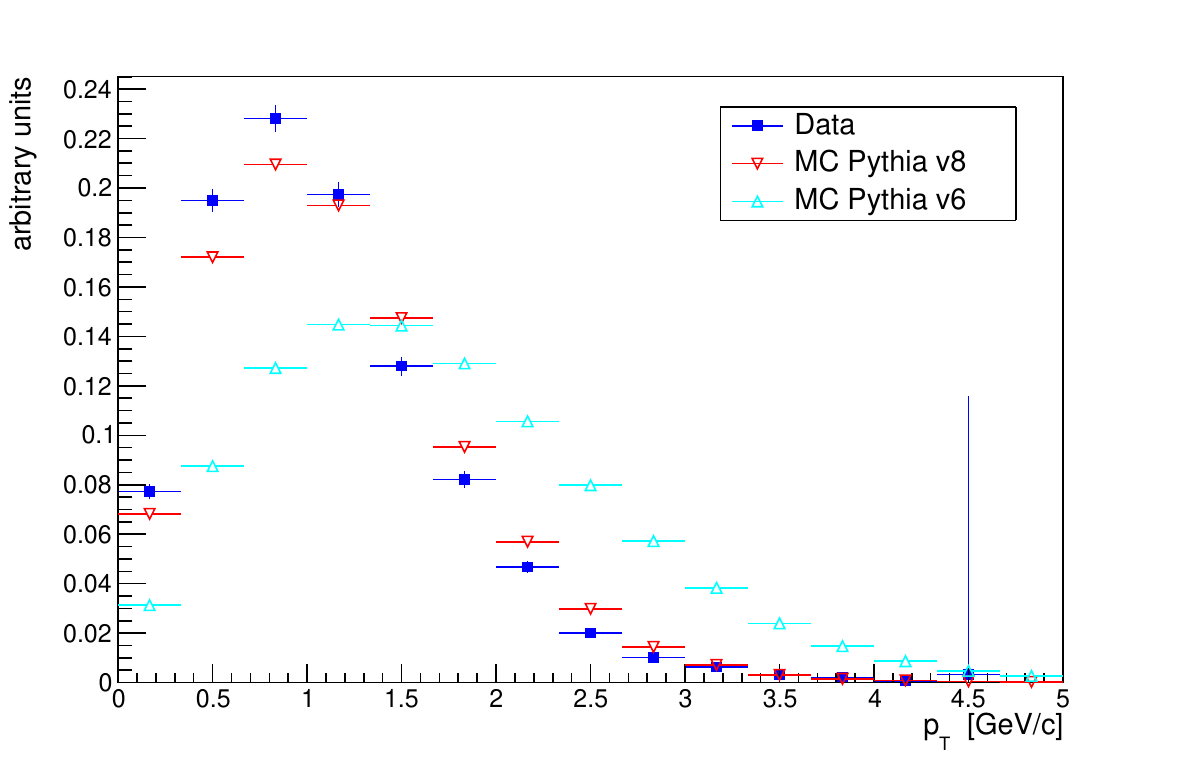}  
\includegraphics[width=\columnwidth]{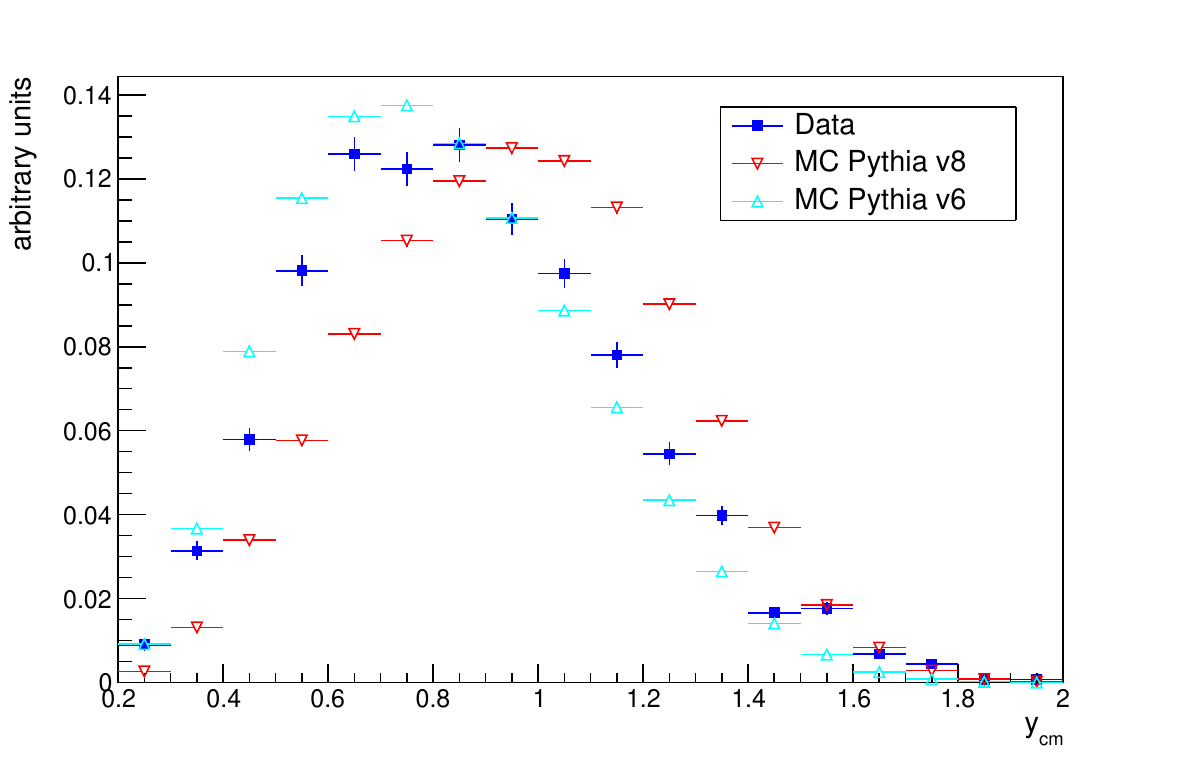}
\caption{Comparing \MC with \data. Number of reconstructed \jpsi as function of its transverse momentum (top) and as function of its rapidity (bottom). }
\label{fig:JpsiPtYReco}
\end{figure}

Pythia v8 represents more closely the data when comparing the number of reconstructed \jpsi as a function of $p_T$, as shown in Figure~\ref{fig:JpsiPtYReco}, top. The rapidity distributions are also very different between Pythia v6 and Pythia v8. In this case, Pythia v6 represents more closely the data as shown in Figure~\ref{fig:JpsiPtYReco}, bottom. To obtain the cross section (see Section~\ref{sec:xsec}), the Pythia v6 \MC was reweighted to follow the Pythia v8 transverse momentum distribution, while the Pythia v8 \MC was reweighted  to follow the Pythia v6 rapidity distribution.

\begin{figure}[ht]
\centering
\includegraphics[width=0.9\columnwidth]{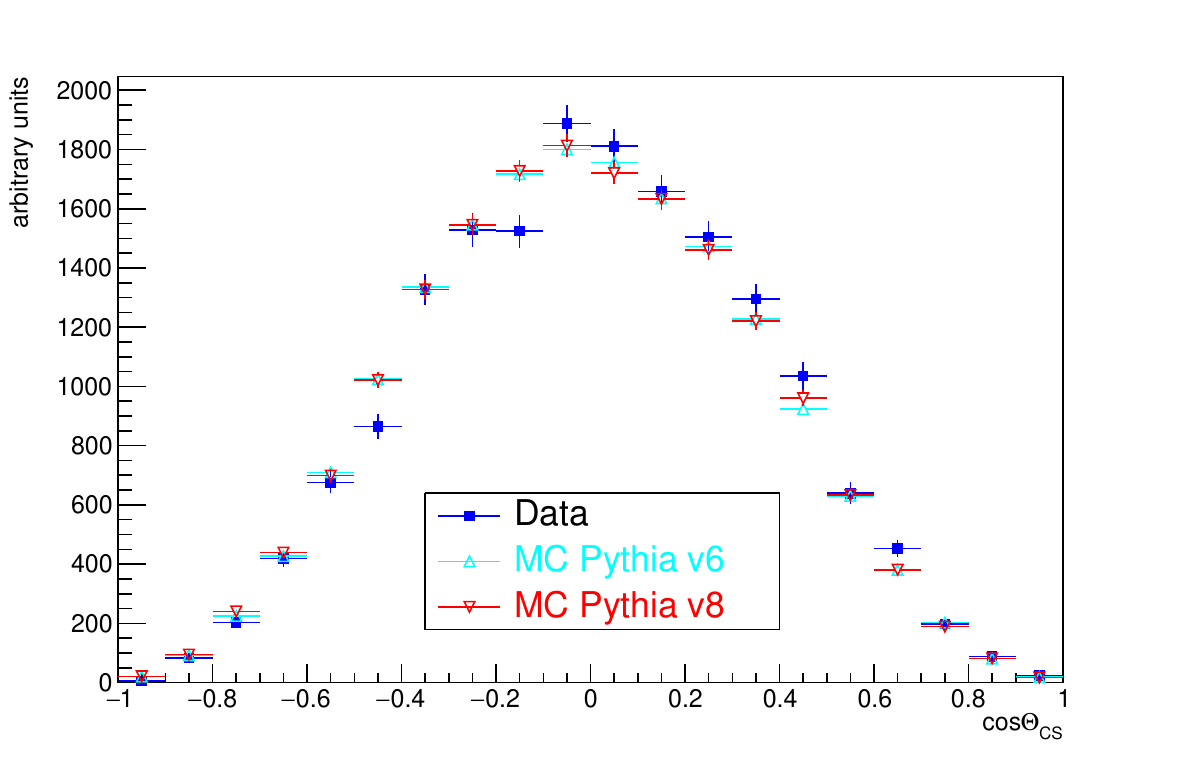}
\caption{Reconstructed \jpsi yields in bins of \cosTheta integrated over $\ycm [0.4 - 1.6]$.}
\label{fig:CSreco}
\end{figure}

In the \MC there is no polarization of the \jpsi, i.e. the \cosTheta distribution is flat. It is however severely affected by the angular acceptance of the experimental setup, shaping the reconstructed distribution, as can be seen in Figure~\ref{fig:CSreco}.

\section{Cross-section as function of rapidity and \cosTheta}
\label{sec:xsec}

To build the differential cross-section distribution as a function of \ycm, the \jpsi signal yield is extracted by fitting the di-muon invariant mass distribution in bins of \ycm, correcting for acceptance $\cal{A}$ and efficiency $\varepsilon$ factors, and resolution effects.

\begin{figure}[ht]
\centering
\includegraphics[width=\columnwidth]{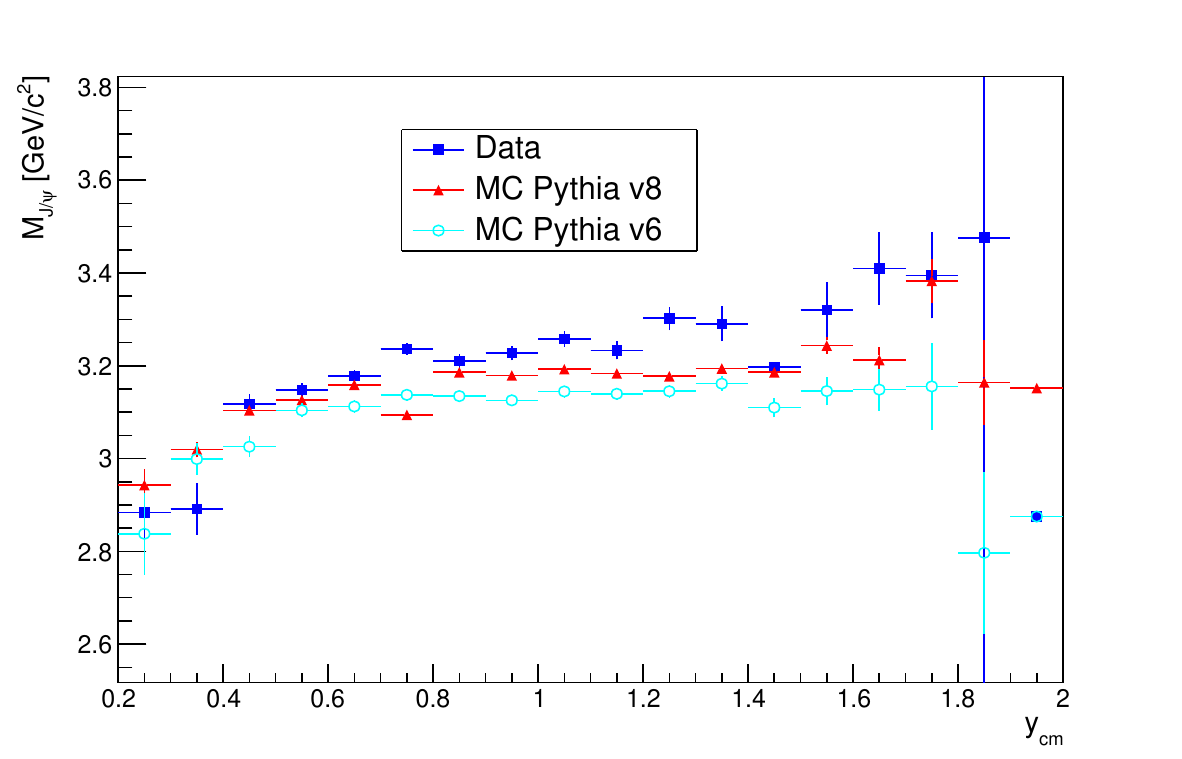}
\caption{Fitted \jpsi mean mass as function of \ycm for \data and \MC. }
\label{fig:fitsys:mass}
\end{figure}

\begin{figure}[ht]
\centering
\includegraphics[width=\columnwidth]{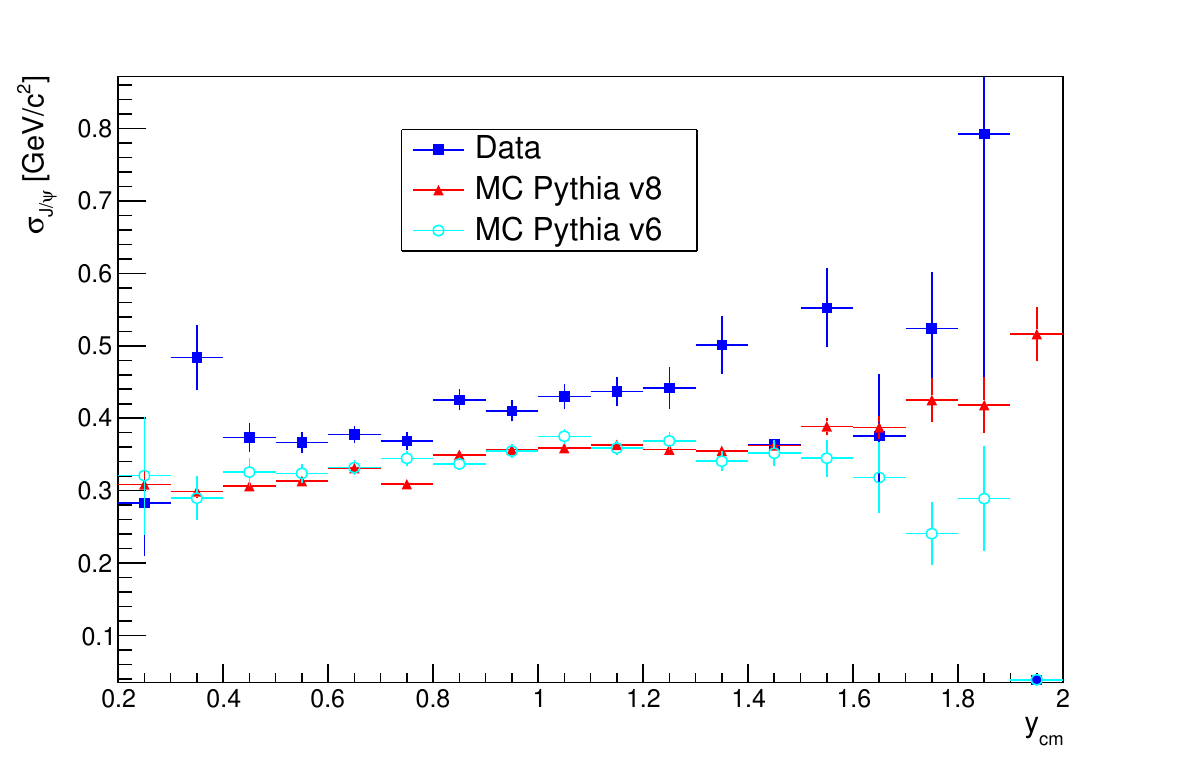}
\caption{Fitted \jpsi mass resolution as function of \ycm for \data and \MC. }
\label{fig:fitsys_sigma}
\end{figure}

\begin{figure}[ht]
\centering
\includegraphics[width=0.8\columnwidth]{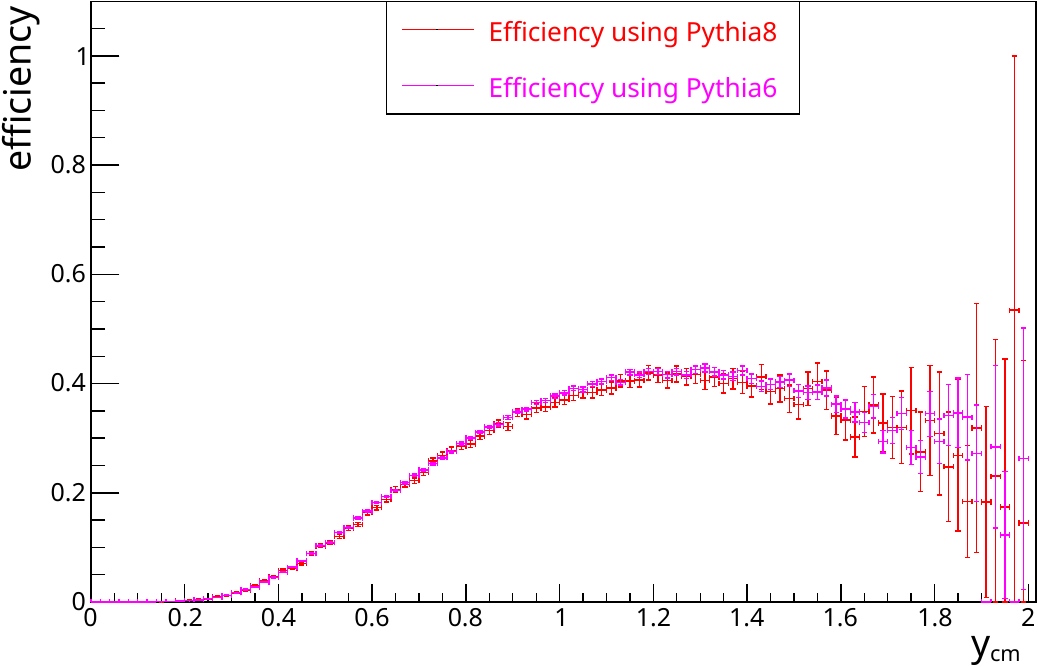}
\caption{Reconstruction acceptance times efficiency ($\cal{A}\times \varepsilon$) obtained from the weighted Pythia v6 (prompt interactions) and Pythia v8 samples .}
\label{fig:jpsieff}
\end{figure}

Figures~\ref{fig:fitsys:mass} and \ref{fig:fitsys_sigma}  show the dependence of the \jpsi mass and the mass resolution  on \ycm.
The product of $\cal{A}\times \varepsilon$ is shown in Figure~\ref{fig:jpsieff} as function of the true rapidity from the two \MC samples produced with Pythia v6 and Pythia v8, weighted to better match the data distributions as described in Section~\ref{sec:recoJpsi}.
Due to the limited angular coverage, $\cal{A}\times \varepsilon$ starts increasing from $\ycm \approx 0.3$ and reaches about $40\%$ at $\ycm \approx 1.0$.

When correcting the \data for acceptance and efficiency, the migration $M(y_{\mathrm{true}},y_{\mathrm{rec}})$ between true and reconstructed \ycm caused by resolution effects  needs to be taken into account.  From the Monte Carlo simulation, we extract for a given interval $\mathrm{d}y_{\mathrm{rec}}$ the distribution of events as function of the true value of rapidity $g(y_{\mathrm{true}})$:
\begin{equation}
g(y_{\mathrm{true}}) = \int M(y_{\mathrm{true}},y_{\mathrm{rec}}) f( y_{\mathrm{rec}})\mathrm{d}y_{\mathrm{rec}}
\end{equation}
The resulting migration matrix is used to correct the data.

\begin{figure}[ht]
\centering
\includegraphics[width=\columnwidth]{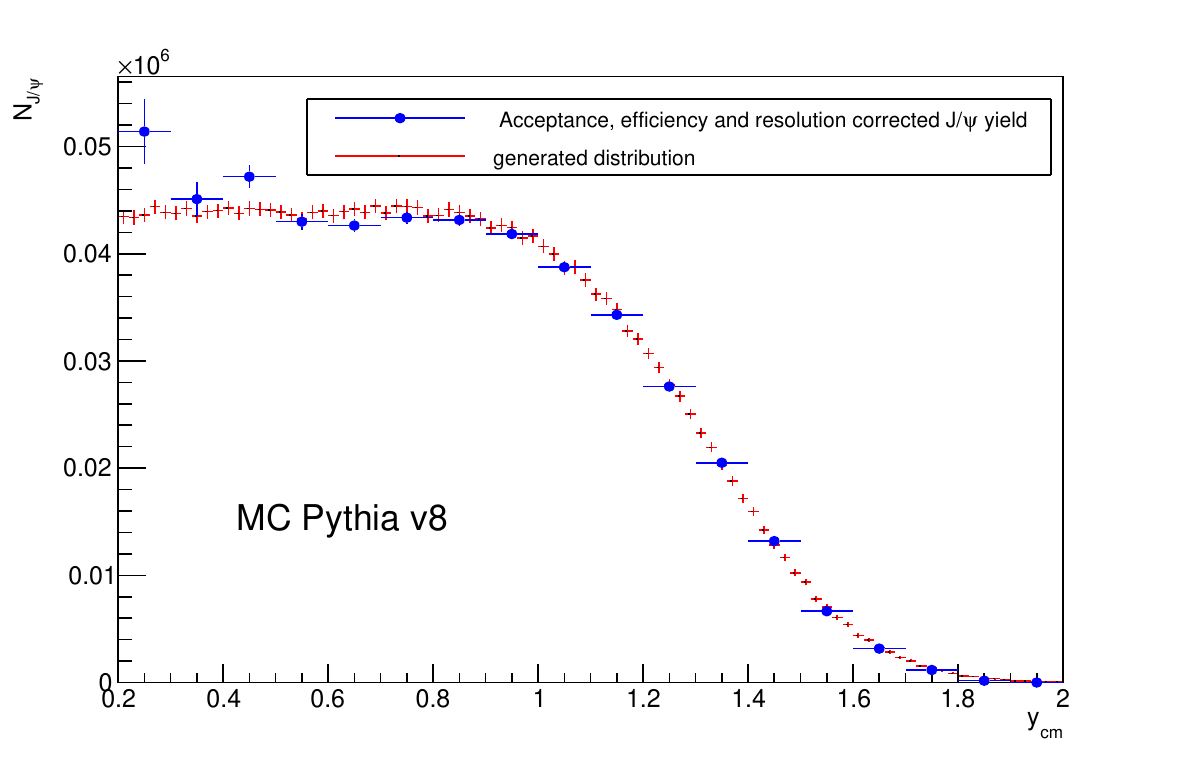}
\includegraphics[width=\columnwidth]{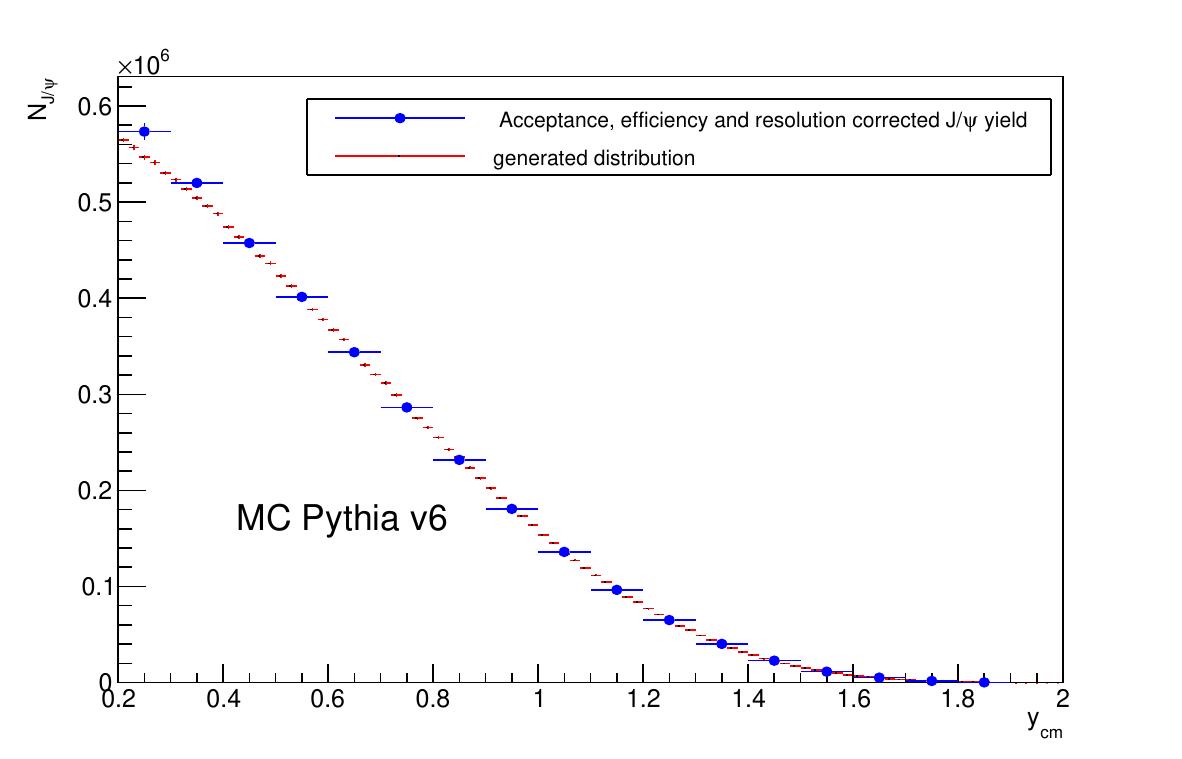}
\caption{Acceptance, efficiency and resolution corrected \jpsi yield (full circle) vs. \ycm in the \MC compared to the generated distribution (dots), Pythia v8 (top) and Pythia v6 (bottom).}
\label{fig:effCorrMC}
\end{figure}

Figure~\ref{fig:effCorrMC} shows the corrected yield as function of \ycm compared to the generated distribution for Pythia v8 (top) and Pythia v6 (bottom). The differences between generated and estimated yields are taken as a systematic error for the final result.

\begin{figure}[ht]
\centering
\includegraphics[width=\columnwidth]{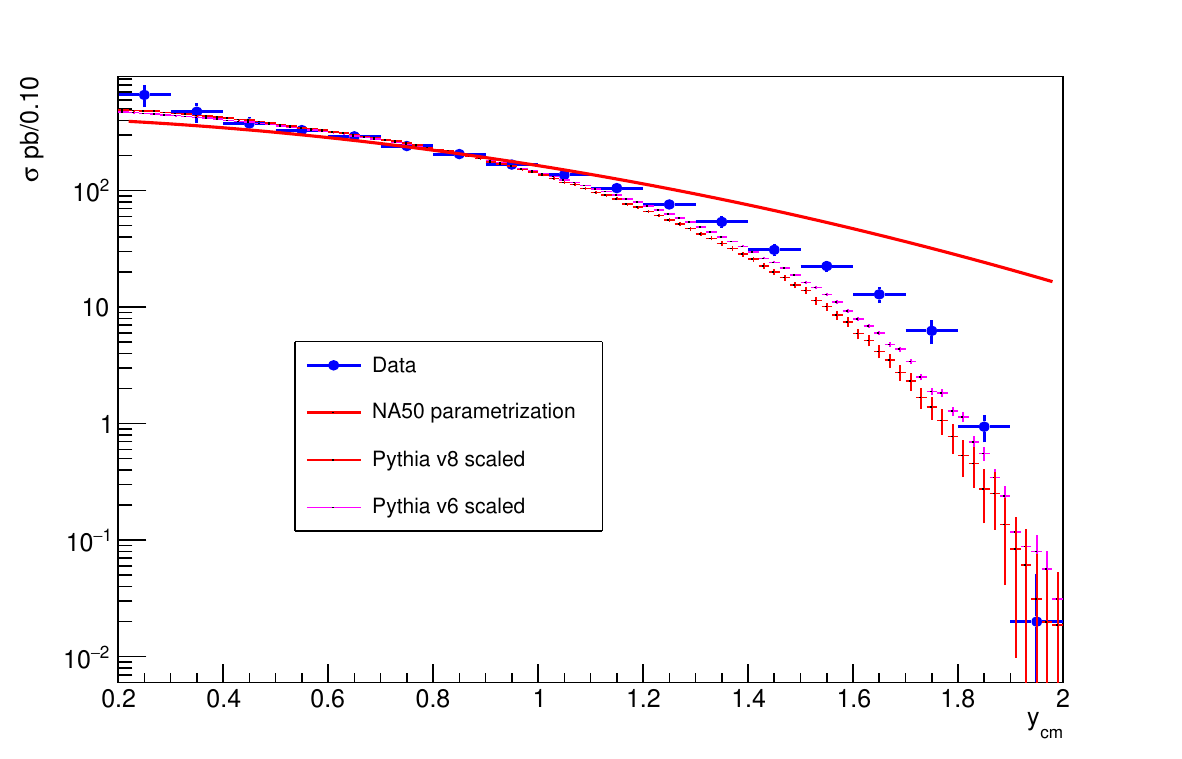}
\caption{Measured \jpsi cross-section as a function of \ycm. The red curve is the extrapolation of the NA50 measurement using their parametrization.}
\label{fig:jpsixsec}
\end{figure}

The results (using the Bukin fit model and the deviations from the results using the Crystal Ball fit model are added to the systematic errors) are shown in Figure~\ref{fig:jpsixsec} and summarized in Table~\ref{tab:jpsixsec}.
The extrapolation of the NA50 measurement~\cite{Alessandro:2006jt} using their parametrization is shown  for the y-dependence.  For the interval with the largest overlap with the NA50 measurement, $0.3 < \ycm < 0.6$, we obtain including the branching ratios of $\jpsi$ decays into muons, $B_{\mumu}$:

$B_{\mumu}\sigma(\jpsi) / A =  (1.18 \pm  0.04 \pm 0.10)~\mathrm{nb}$/nucleon

to be compared to the NA50 (extrapolated using their parametrization) result\cite{Alessandro:2006jt} using a much thinner target of:

$B_{\mumu}\sigma(\jpsi) / A =  (0.99 \pm  0.04)~\mathrm{nb}$/nucleon.

Within the systematic errors, no significant enhancement due to secondary production of \jpsi is observed. An upper limit of possible contributions from secondary collisions of $<32\%$ at 90\% confidence level is obtained from a comparison with NA50. It should be noted that the \muflux measurement extends to very high \ycm, close to the kinematical limit.
\begin{table*}[t]
\centering
\caption{Measured \jpsi cross-section in intervals of \ycm.}
\label{tab:jpsixsec}
\begin{NiceTabular}{|r| c | c | c | }
\hline
Rapidity \ycm &cross-section& statistical  & systematic \\
  Interval &$B_{\mumu}\sigma(\jpsi) / A ( \mathrm{pb/nucleon} )$&error   &  error\\
\hline
$ 0.2 -  0.3$ & $  660.34$ & $ 84.63 $ & $113.7        $   \\
$ 0.3 -  0.4$ & $  473.61$ & $ 35.89 $ & $86.24        $   \\
$ 0.4 -  0.5$ & $  375.56$ & $ 16.97 $ & $49.74        $   \\
$ 0.5 -  0.6$ & $  329.45$ & $ 12.30 $ & $32.63        $   \\
$ 0.6 -  0.7$ & $  291.41$ & $  9.86 $ & $22.38        $   \\
$ 0.7 -  0.8$ & $  240.60$ & $  7.97 $ & $17.09        $   \\
$ 0.8 -  0.9$ & $  205.21$ & $  6.73 $ & $14.56        $   \\
$ 0.9 -  1.0$ & $  166.79$ & $  5.62 $ & $13.15        $   \\
$ 1.0 -  1.1$ & $  135.93$ & $  4.80 $ & $10.38        $   \\
$ 1.1 -  1.2$ & $  104.79$ & $  4.13 $ & $ 9.31        $   \\
$ 1.2 -  1.3$ & $  75.83 $ & $ 3.59 $  & $6.40         $   \\
$ 1.3 -  1.4$ & $  53.78 $ & $ 3.01 $  & $5.27         $   \\
$ 1.4 -  1.5$ & $  30.94 $ & $ 2.14 $  & $3.05         $   \\
$ 1.5 -  1.6$ & $  22.45 $ & $ 1.87 $  & $1.70         $   \\
$ 1.6 -  1.7$ & $  12.84 $ & $ 1.57 $  & $1.26         $   \\
$ 1.7 -  1.8$ & $   6.25 $ & $ 0.98 $  & $1.06         $   \\
$ 1.8 -  1.9$ & $   0.94 $ & $ 0.15 $  & $0.20         $   \\
$ 1.9 -  2.0$ & $   0.02 $ & $ 0.01 $  & $0.03         $   \\
\hline
\end{NiceTabular}
\end{table*}

\begin{figure}[ht]
\centering
\includegraphics[width=\columnwidth]{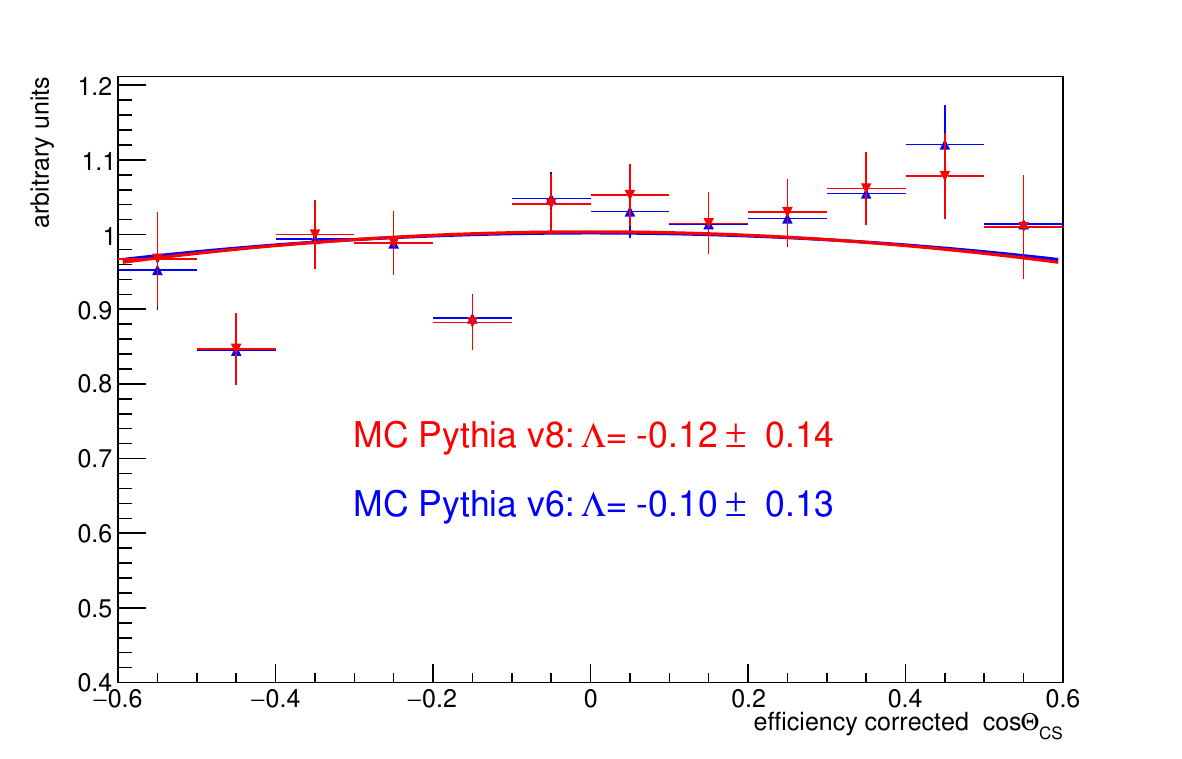}
\caption{Reconstructed and efficiency corrected \jpsi yields in bins of \cosTheta integrated over $\ycm [0.4 - 1.6]$. The data is efficiency corrected either using Pythia v6 (blue) and Pythia v8 (red).
A fit to the two distributions shows no sign of polarization. The polarization parameter $\Lambda$ is compatible with zero.  }
\label{fig:CSaccCorrected}
\end{figure}

The \cosTheta distribution is also corrected for acceptance and efficiency effects, shown in Figure~\ref{fig:CSaccCorrected}. Fitting the data distribution with the function $1 + \Lambda \times (\cosTheta)^2$ yields no significant polarization, $\Lambda = 0.11 \pm 0.14 (\mathrm{stat.}) \pm 0.02 (\mathrm{sys.})$.

\section{Summary}
\label{sec:summary}
A clear signal of \jpsi production is seen in the data of the
SHiP muon flux measurement~\cite{muflux}. The understanding of the rate of \jpsi mesons
in proton fixed target collisions is important for the SHiP
experiment and the design of its magnetic muon shielding, since
they are a source of high \pt muons. $\jpsi$ production
is observed up to high rapidity. The di-muon rate is in reasonable
agreement with Monte Carlo simulations based on Pythia v8. The $\jpsi$ rate observed in the data however is much lower than predicted by the Monte Carlo simulation, which is advantageous for SHiP since  muons with a large $\pt$ are more difficult to deflect by the active muon shield.

We obtain the production cross-section per nucleon $\sigma(\jpsi)/A$ including the branching ratios of $\jpsi$ decays into muons, $B_{\mumu}$:

$B_{\mumu}\sigma(\jpsi) / A =  (1.18 \pm  0.04 \pm 0.10)~\mathrm{nb}$/nucleon 

in the interval with the largest overlap with the NA50 measurement, $0.3 < \ycm < 0.6$. This is  to be compared to the NA50 (extrapolated) result of:

$B_{\mumu}\sigma(\jpsi) / A =  (0.99 \pm  0.04)~\mathrm{nb}$/nucleon. 

Within the systematic errors, no significant enhancement due to secondary production of \jpsi inside the target is observed. The upper limit of possible contributions from secondary collisions is $<32\%$ at 90\% confidence level.

The \cosTheta data distribution shows no significant polarization effect after correcting for acceptance and efficiency factors.
\section*{Acknowledgements}
\label{sec:acknowledgements}
The muon flux measurement took place in 2018 and the analysis presented here was terminated in 2020. Due to subsequent geopolitical and other circumstances the following collaborators, who were members of the SHiP Collaboration in 2018, must be acknowledged:
C.~Ahdida,
A.~Akmete,
A.~Anokhina, 
E.~Atkin, 
T.~Barbe, 
A.~Bagulya,
A.Y.~Berdnikov, 
Y.A.~Berdnikov, \\
S.Bieschke,
A.~Buonaura,
M.~Casolino,
N.~Charitonidis,\\
M.~Chernyavskiy, 
T.~Colin, 
C.~Chatron, 
V.~Dmitrenko, 
O.~Durhan,
E.~Elikkaya,
A.~Etenko, 
O.~Fedin, 
K.~Filippov, 
R.~Froeschl,
G.~Gavrilov, 
V.~Golovtsov, 
D.~Golubkov, 
P.~Gorbounov,
S.~Gorbunov, 
M.~Gorshenkov, 
A.L.~Grandchamp,
V.~Grachev, 
V.~Grichine,
N.~Gruzinskii, 
O.~Id Bahmane, 
Yu.~Guz, 
M.~Huschyn,
M.~Jonker,
D.~Karpenkov, 
M.~Khabibullin, 
E.~Khalikov, \\
A.~Khotyantsev, 
V.~Kim, 
N.~Konovalova, 
I.~Korol'ko, 
A.~Korzenev,
V.~Kostyukhin,
Y.~Kudenko, 
P.~Kurbatov, 
V.~Kurochka, 
L.~Le Mao,  
V.~Maleev, 
R.~Mauny, 
A.~Malinin, 
A.~Mefodev, 
P.~Mermod,
O.~Mineev, 
P.~Moyret, 
S.~Nasybulin,
B.~Obinyakov, 
N.~Okateva, 
B.~Opitz, \\
N.~Owtscharenko,
A.~Petrov, 
D.~Podgrudkov, \\
M.~Prokudin, 
A.B.~Rodrigues~Cavalcante,
T.~Roganova, 
V.~Samsonov, 
E.S.~Savchenko, 
A.~Shakin, 
P.~Shatalov, 
V.~Shevchenko, 
S.~Shirobokov,
A.~Shustov, 
M.~Skorokhvatov, 
S.~Smirnov, 
N.~Starkov, 
M.E.~Stramaglia,
D.~Sukhonos,
P.~Teterin, 
S.~Than~Naing, \\
R.~Tsenov,
S.~Ulin,  
Z.~Uteshev, 
L.~Uvarov,
D.~Valencon, 
K.~Vlasik, 
A.~Volkov, 
R.~Voronkov, 
N.~Wojcicka,
A.~Zelenov.

The SHiP Collaboration wishes to thank National University of Science and Technology (MISIS) who \\ provided the target for this experiment.
We further wish to acknowledge the support from the National Research Foundation of Korea 
with grant numbers of 2018R1A2B2007757, 
2018R1D1A3B07050649, \\
2018R1D1A1B07050701, 2017R1D1A1B03036042, \\
2017R1A6A3A01075752, 2016R1A2B4012302, and \\
2016R1A6A3A11930680, from the FCT - Funda\c{c}\~{a}o para a Ciencia e a Tecnologia of Portugal with 
grant number CERN/FIS-PAR/0030/2017, and from the TAEK of Turkey.

\bibliographystyle{unsrt} 
\bibliography{Biblio}





\onecolumn
\noindent

\centerline{\Large\bf The SHiP Collaboration}
\vspace*{1mm}
\noindent
\begin{center}
A.~Aksoy$^{13}$\orcidlink{0009-0001-3734-2157},
R.~Albanese$^{20,d}$\orcidlink{0000-0003-4586-8068},
K.~Albrecht$^{9}$\orcidlink{0000-0001-5257-3474},
F.~Alessio$^{35}$\orcidlink{0000-0001-5317-1098},
A.~Alexandrov$^{20,d}$\orcidlink{0000-0002-1813-1485},
F.~Aloschi$^{20,d}$\orcidlink{0000-0002-2501-7525},
T.~Arndt$^{14}$\orcidlink{0000-0002-7797-8862},
D.~Arutinov$^{13}$\orcidlink{0000-0001-6150-7030},
A.~Ballarino$^{35}$\orcidlink{0000-0002-1705-4537},
L.~Baudin$^{35}$\orcidlink{0000-0002-1847-6416}
% ...
\end{center}
\vspace*{0.5cm}

{\footnotesize \it

\noindent
% ...
$^{9}$Institut f\"{u}r Physik, Humboldt-Universit\"{a}t zu Berlin, Berlin, Germany\\
% ...
$^{13}$Integrated Computing Architectures (ICA | PGI-4),Forschungszentrum J\"{u}lich GmbH (KFA),  J\"{u}lich , Germany\\
$^{14}$Karlsruhe Institute of Technology, Karlsruhe, Germany \\
% ...
$^{20}$Sezione INFN di Napoli, Napoli, Italy\\
% ...
$^{35}$European Organization for Nuclear Research (CERN), Geneva, Switzerland\\
% ...
$^{d}$Universit\`{a} di Napoli ``Federico II``, Napoli, Italy \\
% ...
}

\end{document}